\documentstyle[twoside,fleqn,espcrc2]{article}
\title{Polarised and Unpolarised Charmonium Production at Higher Orders in $v$}
\author{ Sourendu Gupta and Prakash Mathews
\\
Theory Group,
Tata Institute of Fundamental Research,
Homi Bhabha Road, Bombay 400 005, INDIA
       }

\begin{document}
\begin{abstract}
We study the unpolarised and polarised hadro-production of 
charmonium in non-relativistic QCD (NRQCD) at low transverse momentum, 
including  sufficiently higher orders in the relative velocity, $v$, 
so as to study the ratio of $\chi_{c1}$ and $\chi_{c2}$ production rates.
\end{abstract}
\maketitle

\section{\label{intro}Introduction}

Recent progress in the understanding of cross sections for production of
heavy quarkonium resonances has come through the NRQCD reformulation of 
this problem \cite{caswell}.  A Factorisation approach based on NRQCD 
developed by Bodwin, Braaten and Lepage \cite{bbl} enables the factorisation 
of the production cross sections for a quarkonium $H$ (with momentum $P$) 
into a perturbative part and a non perturbative part---
\begin{equation}
   d\sigma\;=\;{1\over\Phi}{d^3P\over(2\pi)^3 2E_{\scriptscriptstyle P}}
     \sum_{ij} C_{ij}\left\langle{\cal K}_i\Pi(H){\cal 
K}^\dagger_j\right\rangle,
\label{intro.nrqcd}\end{equation}
where $\Phi$ is a flux factor and $\Pi(H)$ denotes the hadronic projection 
operator.   The fermion bilinear operators ${\cal K}_i$ are built out of 
heavy quark fields sandwiching colour and spin matrices and the covariant 
derivative ${\bf D}$.  The labels $i$, $j$ include the colour 
index, spin $S$, orbital angular momentum $L$ (coupling the $N$ covariant 
derivatives), the total angular momentum $J$ and helicity $J_z$. 

The coefficient function $C_{ij}$ is computable in perturbative QCD and 
has an expansion in the strong coupling $\alpha_{\scriptscriptstyle S} 
(m)$ (where $m$ is the mass of heavy quark), whereas the matrix elements 
are non perturbative.  However, in NRQCD, these matrix elements scale as
powers of $v$.
Hence the resulting cross section is an expansion in powers of $\alpha_
{\scriptscriptstyle S} (m)$ and $v$.  Often, higher orders in $v$ involves 
previously neglected colour-octet states of the heavy quark 
pairs.  For charmonium states, a numerical coincidence, $v^2\sim\alpha_
{\scriptscriptstyle S}(m^2)$, makes the higher order terms in $v$ important 
and the double expansion more complicated.  
The above non perturbative matrix elements can be reduced to the diagonal form,
${\cal O}^H_\alpha({}^{2S+1} L_J^N)$ (where $\alpha$ denotes the colour
singlet or octet state) and off-diagonal form, ${\cal P}^H_\alpha
({}^{2S+1}L_J^N,{}^{2S+1}L_J^{N'})$ \cite{bbl}.  These matrix 
elements scale as $v^d$ where $d=3+N+N'+2 (E_d+2M_d)$, $E_d$ and $M_d$ are 
the number of colour electric and magnetic transitions.
This formalism has been successfully applied to large transverse momentum
processes \cite{jpsi}.  Inclusive production cross sections for charmonium 
at low energies, dominated by low transverse momenta, also seem to have a 
good phenomenological description in terms of this approach 
\cite{ours,br,our2}. The spin asymmetries have also been computed both for 
the low transverse momentum processes \cite{pol} and for those with high
transverse momenta \cite{polo}.

It was argued in \cite{our2} that a better understanding of such cross sections,
and asymmetries, can be obtained if the higher order terms in $v$ and
$\alpha_{\scriptscriptstyle S}$ are used. This follows from the fact that 
the total inclusive $J/\psi$ cross sections arise either from direct $J/\psi$
production (which starts at order $\alpha_{\scriptscriptstyle S}v^7$) or
through radiative decays of $\chi_J$ states. $\chi_0$ and $\chi_2$ are first 
produced at order $\alpha_{\scriptscriptstyle S}v^5$, whereas $\chi_1$, which 
has the largest branching fraction into $J/\psi$, is produced only at order
$\alpha_{\scriptscriptstyle S}v^9$.  Further phenomenological problem is to 
explain the $\chi_1/\chi_2$ ratio observed in hadro production 
\cite{beneke,cacci}.  A better understanding of these cross sections 
requires the NRQCD expansion up to order $\alpha_{\scriptscriptstyle S}v^9$. 
The unpolarised and polarisation cross sections are defined as
\begin{equation}
  \sigma \;=\; \sum_{hh'} \sigma(h,h'), \quad 
  \Delta\sigma \;=\; \sum_{hh'} hh'\sigma(h,h'),
\label{intro.asym}\end{equation}
where $h$ and $h'$ are the helicities of the beam and target respectively,
and $\sigma(h,h')$ denotes the cross section for fixed initial helicities.
The difference 
between the polarised and unpolarised case is in the coefficient functions 
denoted by $\tilde C_{ij}$ for the polarised case; the set of non perturbative 
matrix element are the same.  We construct the coefficient functions 
using the ``threshold expansion'' method \cite{bchen} and enumerate the 
non perturbative matrix elements using the spherical tensor method described in 
\cite{upol}.

\section{Cross Sections\label{pol}}

To lowest order in $\alpha_S$ the contributing parton level cross section
are $\bar q q \rightarrow \bar Q Q$ and $g g \rightarrow \bar Q Q$.  The 
hadron level cross section is obtained by multiplying appropriate parton 
luminosities---  
\begin{equation}
   {\cal L}_{ab}\;=\;a(x_1) ~b(x_2),
\label{me.lumin}\end{equation}
where $a$, $b$ runs over quark $q_f$, antiquark $\bar q_f$ or gluon $g$ 
densities depending on the subprocess cross sections. For polarised cross 
sections, $\Delta \sigma$, the corresponding polarised luminosities 
$\Delta {\cal L}_{ab}$ is obtained by replacing $a$, by 
polarised parton densities $\Delta a$.  Data indicates that ${\cal L}_
{\bar qq}\ll{\cal L}_{gg}$ for $\sqrt S\ge20$ GeV and $|\Delta{\cal L}_
{\bar qq}|\ll {\cal L}_{gg}$.  Consequently, the $\bar qq$ channel may 
be neglected for double polarised asymmetries to good precision.

The squared matrix element for the $gg$ process is technically more 
complicated.  The difference between the unpolarised \cite{upol} and 
polarised cases \cite{pol1} lies solely in the flipped sign of the $J=2$ 
part, which
arise in the polarisation sum of initial state gluons. The subprocess 
cross section for the production of a charmonium $H$ can be written as
\begin{equation}\begin{array}{rl}
   \hat\sigma^{H}_{gg}(\hat s) &=
                  {\displaystyle{\pi^3\alpha_s^2\over4m^2}}\delta(\hat s-4 m^2)
		  ~~\sum_d
		\nonumber\\&
                  \left[{\displaystyle{1\over18}}\Theta^{H}_S(d)
                  +{\displaystyle{5\over48}}\Theta^{H}_D(d)
                        +{\displaystyle{3\over16}}\Theta^{H}_F(d)
		  \right],
\end{array}\label{cs.jpsi}\end{equation}
where $d$ runs over the various matrix elements that contribute to the 
charmonium $H$ at order $v^d$.  The subscript $S$, $D$ and $F$ denotes 
the colour singlet, colour octet symmetric and antisymmetric parts 
respectively.  For the polarised case $\hat\sigma^H$ is replaced by 
$\Delta\hat\sigma^H$ and the combination of non perturbative matrix 
elements $\Theta^H$ by $\widetilde\Theta^H$.  For the various charmonium 
states $J/\psi$ and $\chi_J$ the combination of non perturbative matrix 
elements are listed bellow.  The changes for the polarised case is 
mentioned appropriately.  

Direct $J/\psi$ Production
\begin{equation}\begin{array}{rl}
             \Theta^{J/\psi}_D(7)&=
        {\displaystyle{1\over2m^2}}{\cal O}^{J/\psi}_8({}^1S_0^0)
         \\&
       +{\displaystyle{1\over2m^4}}\left[
               3{\cal O}^{J/\psi}_8({}^3P_0^1)
              +{\displaystyle{4\over5}}{\cal O}^{J/\psi}_8({}^3P_2^1)
                                   \right], \\

             \Theta^{J/\psi}_D(9)&=
        {\displaystyle{1\over\sqrt3m^4}}{\cal P}^{J/\psi}_8({}^1S_0^0,{}^1S_0^2)
       +{\displaystyle{1\over\sqrt{15}m^6}}
      \\&
           \biggl[
           {\displaystyle{35\over4}}
                    {\cal P}^{J/\psi}_8({}^3P_0^1,{}^3P_0^3)
          +2 {\cal P}^{J/\psi}_8({}^3P_2^1,{}^3P_2^3)
                                   \biggr], \\

             \Theta^{J/\psi}_F(9)&= \displaystyle{1\over2 m^6} 
               \left[{1\over3} {\cal O}^{J/\psi}_8 ({}^3P^2_1)-
             {2\over5} {\cal O}^{J/\psi}_8 ({}^3P^2_2) \right].\\
\end{array}\label{cs.jpsime}\end{equation}

$\chi_0$ Production

\begin{equation}\begin{array}{rl}
   \Theta^{\chi_0}_S(5)&=
        {\displaystyle{3\over2m^4}}{\cal O}^{\chi_0}_1({}^3P_0^1),\\

   \Theta^{\chi_0}_S(7)&=
        {\displaystyle{7\sqrt5\over4\sqrt3m^6}}
                {\cal P}^{\chi_0}_1({}^3P_0^1,{}^3P_0^3),\\

   \Theta^{\chi_0}_S(9)&=
       {\displaystyle{1\over8m^8}}\biggl[
           {\displaystyle{245\over9}}{\cal O}^{\chi_0}_1({}^3P_0^3)
         \\&
          +{\displaystyle{149\sqrt7\over10\sqrt3}}
               {\cal P}^{\chi_0}_1({}^3P_0^1,{}^3P_0^5)
                                  \biggr]
       +{\displaystyle{2\over5m^4}}{\cal O}^{\chi_0}_1({}^3P_2^1),\\

   \Theta^{\chi_0}_D(9)&=
        {\displaystyle{1\over2m^2}}{\cal O}^{\chi_0}_8({}^1S_0^0)
         \\&
       +{\displaystyle{1\over2m^4}}\left[
               3{\cal O}^{\chi_0}_8({}^3P_0^1)
              +{\displaystyle{4\over5}}{\cal O}^{\chi_0}_8({}^3P_2^1)
                                   \right],\\

   \Theta^{\chi_0}_F(9)&=
           \displaystyle{1\over 6 m^4}{\cal O}^{\chi_0}_8 ({}^1P^1_1)
         \\&
           +\displaystyle{1\over18m^6} \biggl[
                  {\cal O}^{\chi_0}_8 ({}^3S^2_1)
                 +5{\cal O}^{\chi_0}_8 ({}^3D^2_1)
					   \biggr].
\end{array}\label{cs.chi0me}\end{equation}

$\chi_1$ Production
\begin{equation}\begin{array}{rl}
   \Theta^{\chi_1}_S(9)&=
       {\displaystyle{1\over2m^4}}\left[
               3{\cal O}^{\chi_1}_1({}^3P_0^1)
              +{\displaystyle{4\over5}}{\cal O}^{\chi_1}_1({}^3P_2^1)
                                   \right], \\

   \Theta^{\chi_1}_D(9)&=
        {\displaystyle{1\over2m^2}}{\cal O}^{\chi_1}_8({}^1S_0^0)
         \\&
       +{\displaystyle{1\over2m^4}}\left[
               3{\cal O}^{\chi_1}_8({}^3P_0^1)
              +{\displaystyle{4\over5}}{\cal O}^{\chi_1}_8({}^3P_2^1)
                                   \right], \\

   \Theta^{\chi_1}_F(9)&= \displaystyle{1\over 6 m^4}  
        {\cal O}^{\chi_1}_8 ({}^1P^1_1) 
       +\displaystyle{1\over3m^6} \biggl[
        \displaystyle{1\over6}{\cal O}^{\chi_1}_8 ({}^3S^2_1)
      \\&
        +\displaystyle{5\over6}{\cal O}^{\chi_1}_8 ({}^3D^2_1) 
        -\displaystyle{1\over5}{\cal O}_8 ({}^3D^2_2) 
					   \biggr].
\end{array}\label{cs.chi1me}\end{equation}
In eqs.(\ref{cs.jpsime},\ref{cs.chi0me},\ref{cs.chi1me}) the coefficient 
of $J=2$ matrix elements changes sign for polarised cases.  

$\chi_1$ is produced first at order $v^9$. The large branching ratio
for the decay $\chi_1\to J/\psi$ makes this a phenomenologically important
term, and is the main motivation for this work.

$\chi_2$ Production
\begin{equation}\begin{array}{rl}
   \Theta^{\chi_2}_S(5)&=
        {\displaystyle{2\over5m^4}}{\cal O}^{\chi_2}_1({}^3P_2^1),\\

   \Theta^{\chi_2}_S(7)&=
      {\displaystyle{2\over\sqrt{15}m^6}}
                {\cal P}^{\chi_2}_1({}^3P_2^1,{}^3P_2^3),\\

   \Theta^{\chi_2}_S(9)&=
        {\displaystyle{3\over2m^4}}{\cal O}^{\chi_2}_1({}^3P_0^1)
       +{\displaystyle{1\over75m^8}}\biggl[
           {\displaystyle{262\over9}}{\cal O}^{\chi_2}_1({}^3P_2^3)
         \\&
          +{\displaystyle{141\sqrt3\over2\sqrt7}}
                  {\cal P}^{\chi_2}_1({}^3P_2^1,{}^3P_2^5)
                                  \biggr],\\

   \Theta^{\chi_2}_D(9)&=
        {\displaystyle{1\over2m^2}}{\cal O}^{\chi_2}_8({}^1S_0^0)
         \\&
       +{\displaystyle{1\over2m^4}}\left[
               3{\cal O}^{\chi_2}_8({}^3P_0^1)
              +{\displaystyle{4\over5}}{\cal O}^{\chi_2}_8({}^3P_2^1)
                                   \right], \\

   \Theta^{\chi_2}_F(9)&= \displaystyle{1\over 6 m^4} 
        {\cal O}^{\chi_2}_8 ({}^1P^1_1) 
         \\&
        + \displaystyle{1\over3m^6} \biggl[
        \displaystyle{1\over6}{\cal O}^{\chi_2}_8 ({}^3S^2_1) 
        +\displaystyle{5\over6}{\cal O}^{\chi_2}_8 ({}^3D^2_1)
       \\&
        -\displaystyle{1\over5}{\cal O}^{\chi_2}_8 ({}^3D^2_2) 
        +\displaystyle{2\over7}{\cal O}^{\chi_2}_8 ({}^3D^2_3) 
					   \biggr].
\end{array}\label{cs.chi2me}\end{equation}
In the combination ${\Theta}^{\chi_2}_S (9)$, the coefficient of ${\cal O}^
{\chi_2}_1 ({}^3 P^3_2)$ is different for the polarised case, both in magnitude 
as well as sign. This is because the matrix element arises also from sources 
other than the polarisation sum that contribute to the $J=2$ part. To obtain
the polarised expression replace the coefficient $262$ by $- 238$.  The other
$J=2$ terms change sign as expected.  The $J=3$ term in the $F$ colour
amplitude is also derived from the $J=2$ part of the polarisation sum, 
and hence flips sign compared to the unpolarised case.

\section{Discussion\label{disc}}

In spite of the large number of unknown non perturbative matrix elements 
in the final results, it is possible to make several quantitative and 
qualitative comments about the polarisation asymmetries by making use of 
heavy quark spin symmetry and scaling arguments developed in \cite{pol1}. 
This scaling argument allows us to make rough estimates.  Neglecting 
possible logarithms of $m$ and $v$, dimensional argument can be used to 
write
\begin{equation}
   \langle{\cal K}_i\Pi(H){\cal K}^\dagger_j\rangle\;=\;
       R_H Y_{ij} \Lambda^{D_{ij}} v^d,
\label{disc.dimen}\end{equation}
where $D_{ij}$ is the mass dimension of the operator, $d$ is the
velocity scaling exponent in NRQCD, $\Lambda$ is
the cutoff scale below which NRQCD is defined ($\Lambda\sim m$), and
$Y_{ij}$ and $R_H$ are dimensionless numbers. 
$R_H$ contains the irreducible minimum non perturbative information. 

\begin{figure}
\vskip7truecm
\includegraphics{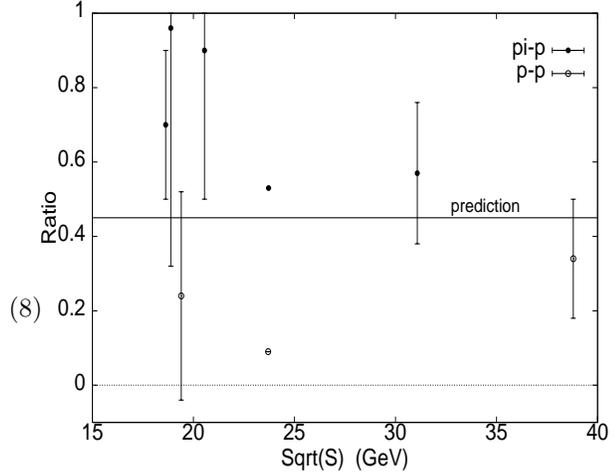}
\caption[dummy]{The ratio $\sigma(\chi_1)/\sigma(\chi_2)$ measured in
$p N$ and $\pi N$ has been plotted as a function of $\sqrt S$.  The data
has be compiled in \cite{beneke}.  Preliminary result presented by E771 
at the Moriond conference is $0.45\pm0.2$ \cite{cacci} and in agreement 
with our estimates. 
 }
\label{ratio}\end{figure}

Assuming the constancy of $R_H$ \cite{pol1} and using heavy-quark symmetry, 
we find that the non perturbative matrix elements contribute approximately 
$12 R_\chi m^2 v^9$ to the $\chi_1$ cross section and about $(1+v^2+12 v^4)
R_\chi m^2 v^5$ to the $\chi_2$ cross sections. Then we expect
\begin{equation}
   {\sigma(\chi_1)\over\sigma(\chi_2)}\;\approx\;
        {12v^4\over1+v^2+12v^4}\;=\;0.45,
\label{disc.rat}\end{equation}
independent of $\sqrt S$. This estimate
is in reasonable agreement with
the measured values in proton-nucleon collisions (see fig.~1)--- $0.34\pm0.16$ at $\sqrt
S=38.8$ GeV \cite{e771} and $0.24\pm0.28$ at $\sqrt S=19.4$ \cite{e673}. The
measurements are also compatible with a lack of $\sqrt S$ dependence. 
Estimate of ${\cal O} (\alpha^3_{\scriptscriptstyle S})$ effects in \cite{br} 
was used to show that the $\chi_1/\chi_2$ ratio could be about $0.3$. In
NRQCD this ratio cannot depend on the beam hadron. It turns out that the
estimate in eq.\ (\ref{disc.rat}) is not very far from the recently measured
value in pion-nucleon collisions--- $0.57\pm0.19$ at $\sqrt S=31.1$ GeV
\cite{e706}. However, the experimental situation certainly needs
clarification.

A straightforward application of the NRQCD scaling laws would lead us
to the conclusion that the asymmetries for $pp\to\chi_{0,2}$
are given by
\begin{equation}
   A_{pp}^{\chi_0}\;\approx\;
   -A_{pp}^{\chi_2}\;\approx\;
       {\Delta{\cal L}_{gg}\over{\cal L}_{gg}}+{\cal O}(v^4),
\label{disc.ppchi02}\end{equation}
where we have neglected the contribution of the $\bar qq$ channel. The
asymmetry for $\chi_1$ production is
\begin{equation}
   A_{pp}^{\chi_1}=
     {{\widetilde\Theta}_S^{\chi_1}(9) + {\widetilde\Theta}_D^{\chi_1}(9)
                                   + {\widetilde\Theta}_F^{\chi_1}(9)
     \over
     \Theta_S^{\chi_1}(9) + \Theta_D^{\chi_1}(9) + \Theta_F^{\chi_1}(9)}
       \left[{\Delta{\cal L}_{gg}\over{\cal L}_{gg}}\right].
\label{disc.ppchip}\end{equation}
The ratio of matrix elements can be estimated using heavy quark spin
symmetry and the scaling relations in eq.\ (\ref{disc.dimen}).
${\widetilde\Theta}_D^{\chi_J}(9)$ vanishes in this approximation and
the terms ${\widetilde\Theta}_{S,F}^{\chi_J}(9)$ come with opposite
signs. The numerator is positive but small and we expect---
\begin{equation}
   A_{pp}^{\chi_1}\;\approx\;
       0.2\,{\Delta{\cal L}_{gg}\over{\cal L}_{gg}}.
\label{disc.ppchi1}\end{equation}
The $\bar qq$ channel remains negligible even at $\sqrt S=500$ GeV.

The $J/\psi$ asymmetry seems to be enormously complicated because of
the radiative decays of the $\chi$ states. However, a
major simplification occurs because of the near vanishing asymmetry
in direct $J/\psi$ production. 
Thus the asymmetry comes entirely from the 20--40\% of the cross section
due to $\chi$ decays. Taking into account the ratios of the production
cross sections of $\chi$ and the branching fractions for their decays
into $J/\psi$, we find that the $\chi_1$ and $\chi_2$ states contribute
equally to $J/\psi$. Hence the $J/\psi$ polarisation asymmetry is
expected to be approximately
\begin{equation}
   A_{pp}^{J/\psi}\;\approx\; -(0.15\pm0.05)
       {\Delta{\cal L}_{gg}\over{\cal L}_{gg}}.
\label{disc.psi}\end{equation}
We summarise the predictions made on the basis of the NRQCD scaling in
eq.\ (\ref{disc.dimen}) and the assumption of $R_H$ depends only on
the hadron $H$---
\begin{equation}\begin{array}{rl}
   -A_{pp}^{J/\psi}  \approx  
   A_{pp}^{\chi_1}  <  
   A_{pp}^{\chi_0}  =  
   -A_{pp}^{\chi_2}  =  
       {\Delta{\cal L}_{gg}\over{\cal L}_{gg}}.
\end{array}\label{disc.nrqcd}\end{equation}

In conclusion, low energy double polarised asymmetries are a good test-bed
for understanding the origin of all observed systematics in fixed target
hadro-production of charmonium. The high order computations
presented here provides a set of processes which can be used to test
aspects of NRQCD factorisation and scaling. 

\section*{Acknowledgements}
One of us PM, would like to thank Prof. S Narison for the invitation to attend
the QCD Euroconference 97 and for the opportunity to present this work.

\end{document}